\documentstyle[12pt,aps]{revtex} 
\input{psfig.sty}

\begin{document}
\title{Lattice-dynamical calculation of phonon scattering at a disordered interface}
\author{G. Fagas, A.G. Kozorezov, C.J. Lambert, J.K. Wigmore}
\address{School of Physics and Chemistry, Lancaster University, LA1 4YB, UK}
\maketitle

\begin{abstract}
For an fcc crystal with central force
interactions and separately for a scalar model on a square lattice
, we compute exactly the phonon transmission coefficient $T(\omega)$
through a disordered planar interface between two identical semi - infinite
leads. At high frequencies $T(\omega)$ exhibits a strong frequency
dependence which is determined by the correlation length of the disorder.
\end{abstract}

\newpage
The problem of energy transfer between two solids
occurs in a wide range of physical structures, including
semiconductor quantum wells and superlattices.
Early studies have assumed a perfect solid-solid interface and as
a first approximation, an elastic continuum model accounting for acoustic
mismatch has been employed to quantify heat flow through the interface \cite{Little}.
Such a continuum theory describes generic low-frequency properties, but e.g. to
account for experimental observations of the Kapitza thermal
conductance between solids \cite{Swartz}
at high temperatures, a lattice-dynamical approach \cite{Maris}
is required, which incorporates important parameters such as the phonon dispersion.

The theory of reference \cite{Maris} provides a comprehensive lattice-dynamical
description of perfect solid-solid interface.
In this paper we discuss one extra mechanism which might play a significant role
in energy transfer, namely phonon scattering due to interfacial disorder.
Breaking the translational invariance at the interface gives rise
to the so-called scattering-mediated phonon transmission(reflection) \cite{WW}
, for which calculations based on pertubation theory
predict a strong frequency dependence with a power-law crossover
determined by the disorder characteristics\cite{Alex}.
We study these effects by analysing a lattice-dynamical
model and calculate exactly the phonon transmission coefficient
as a function of frequency. The disorder is introduced as a
correlated or uncorrelated random variation of masses $m_{\bf j}$
along the lattice sites $\bf j$ of the interface layer.

\begin{figure}
\centerline{\psfig{figure=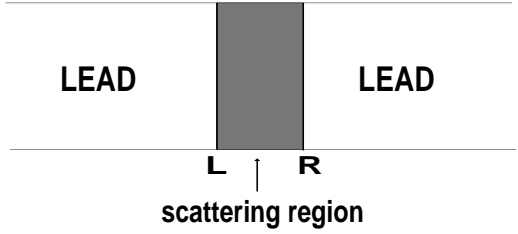,height=3.cm,width=7.5cm}}
\caption{A typical scattering geometry, where L(R) refer to sites
on the faces of left(right) lead.}
\label{struc}
\end{figure}

We formulate the problem of phonon transmission
across a disordered solid-solid interface by considering two
identical semi-infinite leads attached to a 'scattering region'.
The two semi-infinite perfect crystals are envisaged as waveguides
for incident and scattered phonons. The scattering region(disordered interface)
consists of a single atomic plane where interactions with the impurities take place.
A typical geometry is shown in Fig.\ref{struc}.
\newpage

As a model we take an fcc lattice of masses $m_{\bf j}$ coupled to each other
by central force springs of stiffness $\kappa$. As  a second example we also consider
a square lattice described by a Born model with a single degree of freedom per
site. In the latter, phonon polarization is neglected.
In the harmonic approximation the lattice-dynamics in both cases 
is described by the linearized
equations of motion of atomic displacements ${\bf u_{j}}$

\begin{equation}
m_{{\bf j}}\ddot{u}_{
{\bf j}}^{\alpha }=-\sum_{\beta,{\bf i}}K_{{\bf ij}}^{\alpha \beta }u_{{\bf i}
}^{\beta },
\label{motion}
\end{equation}

where $K_{{\bf ij}}^{\alpha \beta }=\kappa(l_{{\bf ij}}^{\alpha }l_{%
{\bf ij}}^{\beta }-4\delta _{{\bf ij}}\delta ^{\alpha \beta })$, $l_{{\bf ij}%
}^{\alpha }={\bf (j-i)}^{\alpha }/{\bf \left| j-i\right| }$, and ${\bf (j-i)}$
is taken from the first coordination sphere. The indices $\alpha,\beta$ run
through $x,y,z$ when phonons can be polarized and only through $z$ when they
are treated as scalars.

In the semi-infinite leads all atoms are of unit mass, whereas
the masses at the interface have mean value $\langle m_{j} \rangle=1$
and variance $\sigma$ \cite{var}.
To generate an ensemble of N positive masses, we first introduce the following
random numbers $\chi_{j} = \sum_{k} (a_{k}\cos(kj) + b_{k}\sin(kj))$
where $a_{k}$ and $b_{k}$ are Gaussian random numbers with zero mean value
and $\langle a_{k}a_{k^\prime}\rangle = \langle b_{k}b_{k^\prime}\rangle =
\delta_{kk^\prime}e^{-\frac{k^2 \xi^2}{4}}$ and j labels
a mass along the interface. Starting from these correlated
quantities with correlation length $\xi$, which may assume positive
and negative values, we define the non-negative quantities $\tilde m_{j}$
and $\hat m_{j}$ via the relations: $\tilde m_{j} = \chi_{j}-\chi_{j}^{min}$,
$\hat m_{j} = \sigma \frac{\tilde m_{j}}
{\sqrt{\langle \tilde m_{j}^2 \rangle-\langle \tilde m_{j} \rangle^2}}$.
Finally the set of correlated mass $\{m_{j}\}$ used in the simulation
is defined by $m_{j} = 1 - \hat m_{j} + \langle \hat m_{j} \rangle$.
In what follows, to isolate the effect of correlations, we compare the transmission
coefficient of such an interface, with that of an uncorrelated interface obtained
by randomly 'shuffling' the above set.

To calculate the overall phonon transmittance
we compute the unitary scattering matrix $S(\omega)$ at a fixed frequency using
the Landauer-B\"{u}ttiker formalism \cite{Landauer}.
In addition to various tests on the numerical code, such as the calculation of the
density of states or of isotropic scattering for a single mass defect,
the unitarity of $S(\omega)$
is checked in all calculations. $SS^{\dagger}=1$ reflects the
conservation of flux. The $S$-matrix is extracted
from the Green's function(Fig.\ref{struc})

\begin{equation}
G(\omega)=\left(\begin{array}{cc} G(L,L;\omega) & G(L,R;\omega)\\
G(R,L;\omega) & G(R,R;\omega)\end{array}\right)
\label{Green}
\end{equation}

\begin{equation}
G(\omega)=(1-g(\omega)D_{s})^{-1}g(\omega),
\label{Dyson}
\end{equation}

where $g(\omega)$ is the Green's function for the leads
calculated using an algorithm developed in \cite{Stefano}.
$D_{s}$ is the dynamical matrix $D=M^{-1/2}KM^{-1/2}$,
$M_{{\bf ij}}^{\alpha \beta }=m_{{\bf i}}\delta ^{\alpha \beta }\delta _{{\bf ij}}$,
which includes all the couplings between atoms on the interface and between
the scattering region and the leads.
Eq.\ref{Dyson} is Dyson's equation written in a convenient
form.

\begin{figure}
\centerline{\psfig{figure=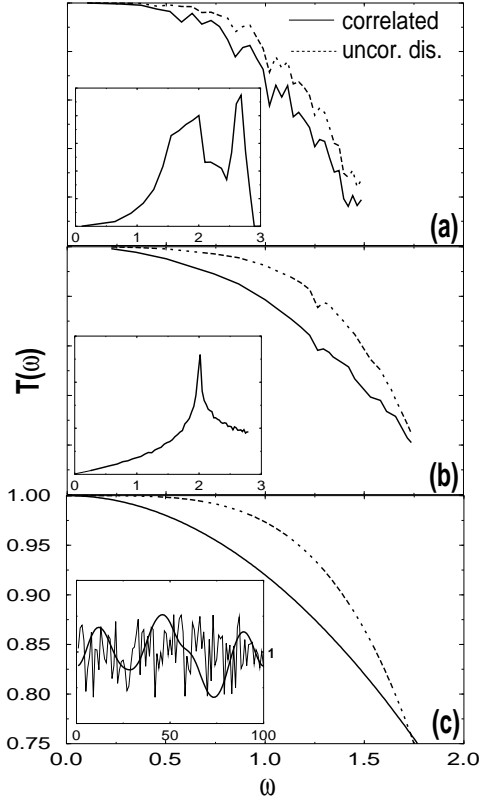,height=11.cm,width=7.5cm}}
\caption{Phonon transmission coefficient for the (a)fcc model($\xi=5.65$), (b)square
scalar model($\xi =20$), (c)estimates based on pertubation theory. In the (a) and (b) insets
the corresponding normalised DOS versus $\omega$ is plotted. Frequency is measured
in units of $\sqrt{\kappa/m}$. Inset (c) shows a single realization
of $m_j$ versus j for a line of correlated and uncorrelated
masses($\sigma=0.3$ in all plots).}
\label{resul}
\end{figure}

In Fig.\ref{resul}(a),(b) the frequency dependence of the
overall transmission coefficient $T(\omega)$ averaged over 10 disorder
realizations of the interface masses
for the fcc vectorial Eq.\ref{motion} and over
50 for the square lattice is shown. The lateral width of the
interface is 29 and 100 in units of the nearest-neighbour spacing.
The reason that we used the simplified scalar model is apparent.
The computational time is quite large if we include phonon
polarization because of the increase in the degrees of freedom.
This prevents us from considering wider structures and eliminiting
finite-size effects evident in the stucture of the plotted
normalised to unity transmittance in Fig.\ref{resul}(a).
In Fig.\ref{resul}(c) the qualitative behaviour
of $T(\omega)$ based on the results of pertubation theory \cite{Alex}
for both correlated and uncorrelated non-ideal interface layer is shown.
These estimates are in good qualitative agreement with the results of the
lattice-dynamical calculations.

Figure 2 shows that a disordered interface of
just a single atomic plane gives rise to strong scattering-controlled
phonon transmission. The overall phonon transmission
coefficient(reflection, $R(\omega) = 1 - T(\omega)$) exhibits strong frequency
dependence with increasing incident phonon frequency. We also 
clearly demonstrate that the precise frequency dependence is determined
by the characteristics of disorder. In particular, there exists a
pronounced difference between the correlated and the uncorrelated
disorder configuration. The exact $T(\omega)$
exhibits much slower frequency dependence for correlated distribution
of masses $m_j$ on the plane of the solid-solid interface than
for uncorrelated configuration obtained by shuffling the same set
$\lbrace m_{j} \rbrace$. Such an effect originates from the
restrictions in the phase volume available for the scattered states
due to the correlation-induced finite width of the disordered
spectral distribution. For the future it would be interesting
to simulate a finite thickness interface between dissimilar solids.

\end{document}